\documentclass{article}  
\usepackage{CJKutf8}
\usepackage{amsmath,amssymb,amsfonts,amsthm} 
\usepackage{graphics}                 
\usepackage{color}                    
\usepackage{hyperref}                 
\usepackage{comment} 			
\usepackage {enumitem}

\pdfoutput=1
\newtheorem{theorem}{定理}[section]

\newtheorem{lemma}[theorem]{引理}

\newtheorem{definition}[theorem]{定义}

\newtheorem{example}{例}[section]

\date{\small\it 2023年8月24日}

\title{叠加体 -- 一种非逻辑的计算模型      
\footnote{感谢我妻子对我的一贯支持。}}

\author{ 熊楚渝 \\
{\small 独立研究员} \\
{\small Email: chuyux99@gmail.com}
}

\begin{document}
\begin{CJK}{UTF8}{gbsn}

\maketitle
\begin{abstract}
我们一直在为寻求超越计算复杂性努力，在此过程中，我们逐渐意识到经典计算的困境，进而意识到叠加体可能是一个解决的方向。叠加体是这样一种模型，它由若干布尔函数的变量和函数值相互复馈形成。叠加体的组成部分是复馈函数，这是经典逻辑可以完全严格描述的，是经典计算可以计算的，但是叠加体整体是一个非逻辑的实体，经典计算不可能对其作完全计算。我们在本文中提出叠加体的概念，并且讨论它的基本性质。我们发现，叠加体+解叠器将形成一种新型的计算模型，其能力可以超越图灵计算。我们设想，这种新型模型将可以帮助在智能体中实现这些功能：一种全新的编程方式和全新的学习方式，内生感受，类比和联想，形成理解，产生能动，参与主观的形成，以及更多。我们还将最初步的讨论如何实现叠加体。
\end{abstract}


{\small\it 才以用而日生，思以引而不竭。
\hspace{270pt} — 王夫之} \\
{\small\it 生命的力量，尤其是心灵的威力，就在于它本身设立矛盾，忍受矛盾，克服矛盾。
\hspace{70pt} — 黑格尔} \\

\vskip 0.3cm

\section{导言} 
我们一直在思考建立超越计算复杂性的智能体\cite{exceeding}。根据\cite{russell}的定义，智能体是一个通过它的感受器从其环境中感知并且通过它的行动器作用于其环境的东西。我们发现，智能体的主观和能动，是提升智能体的能力的关键\cite{subjectivity}。如果智能体具备良好的主观和积极的能动，就可以极大提高智能体的能力。如果我们想建立超越计算复杂性的智能体，就更必须让智能体具备良好的主观和能动。

然而，我们也发现，如果一个智能体仅采用经典计算（图灵计算）的方法，很难实现真正意义下的能动\cite{subjectivity}。这是否意味我们需要发展出一种不同于经典计算的计算，才有可能让智能体具备真正意义下的能动。但是，丘奇-图灵论题说，任何计算都等价于图灵机，怎么可能发展出不同于经典计算的计算？这是一个困境。

在试图克服这样的困境时，我们了解到Findlay, Koch和Toroni等人提出的一个简单模型（我们称为PQR线路）\cite{findlay}。他们的这个简单模型刺激了我们，促使我们深入追问究竟什么是计算，重新思考关于计算的各个方面。在这个方向做了长期思考后，我们萌生了叠加体的思路，因此提出了叠加体--一种新型的计算模型。

简单地说，叠加体是多个布尔函数叠加在一起形成的模型，因此形成了函数的变量和函数值的交叉重叠（术语是复馈）。函数的复馈就形成了多个函数叠加到一起的局面。非常肯定的，这是一种看起来相当混乱的局面，是经典逻辑不能良好描述的情况，是非逻辑性的，也是经典计算不能完全计算的，是一种全新的构型。面对叠加体，面对这种混乱局面，要想回归经典逻辑，就需要从外部施加一种针对复馈的次序，使得可以局部地解脱这种混乱局面，恢复逻辑性。我们把这样的过程称为解叠，而执行解叠的就是解叠器。这样，叠加体+解叠器事实上形成了一种新型的计算模型。在初步探索这个新型的计算模型后，结果令我们感到相当惊喜：这个模型很有潜力，很可能可以帮助我们让智能体具备能动，以及更多。因此，我们在本文中把这个计算模型写下来。

本文中，我们将讨论叠加体+解叠器这个计算模型，以及它的一些基本性质。我们将显示，这种计算模型，为我们提供一种新的编程方式和新的学习方式。我们还将设想，用这个计算模型在智能体中实现若干一些重要功能，如：内生感受，产生能动，形成理解，形成类比和联想，参与智能体的主观，等等。如果使用图灵计算，这些事情是非常难以实现的。这仅是探索叠加体这个模型的最初步，我们将持续研究这个模型的各个方面，包括应用和实现。

本文的安排如下：在2中，我们介绍PQR线路，这是一个来自于意识研究的模型，也是一个非常简单的叠加体。在3中，我们提出叠加和叠加体，以及解叠和解叠器。在4中，我们讨论思绪空间和内蕴函数。在5中，我们设想一些潜在的叠加体的应用，并且初步讨论如何实现叠加体。在6中，我们做一些讨论。

\section{PQR线路} 
PQR线路是Findlay, Koch和Toroni等人提出的\cite{findlay}。若干年前Toroni等人就提出了著名的信息整合理论。信息整合是指：一组信息互为因果，非单一方向影响而是相互影响，进而紧密联系形成一个整体。在为该理论做论证的时候，他们提出了一个非常简单的模型来说明信息整合的理念，因为这个模型使用了3个节点P，Q和R，我们称这个模型为PQR线路，如下图所示。

\begin{center}
\begin{picture}(300,175)(0,0)
\put(38,0){\resizebox{7 cm}{!}{\includegraphics{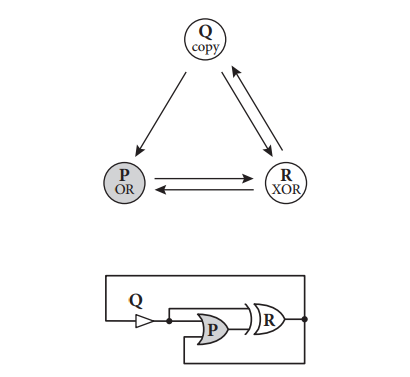}}}
\end{picture}

{图 1. PQR线路，摘自于Koch\cite{koch}} 
\end{center}

可以很清楚看到，这个线路其实就是几个布尔线路的叠加形成的。所谓的叠加，就是在节点处，两个以上的布尔线路复用了该节点。 
我们还可以用布尔函数来表达这个模型。一共有3个布尔线路，因此也就有3个布尔函数。
\begin{align}
& P =f_1 (P,Q,R)=Q \lor R       \label{eq:p}\\
& Q = f_2 (P,Q,R)=R        \label{eq:q} \\    
& R = f_3 (P,Q,R)=P \oplus Q   \label{eq:r}
\end{align}

注意到，如上表达式~\eqref{eq:p}，~\eqref{eq:q}，~\eqref{eq:r}中的每一个函数都是简单的布尔函数，单独来讲，没有什么特别。然而，如果把它们全部放在一起，就可以看到，其特别之处在于，这些变量都是自变量，也是函数值，而且是同时为自变量和函数值。所谓的互为因果，就是这样的。因此，它们就形成了叠加。我们可以把PQR线路看成为3个非常简单的布尔线路叠加在一起而形成的一种特殊的线路。这个线路中，并没有传统的输入节点，也没有传统的输出节点，但是，所有的节点都既是输入节点，也是输出节点。需要注意到，这样的线路和所谓的前馈线路非常不同，也和所谓的循环网络非常不同。

Toroni等人提出了PQR模型，他们的主要论点是，这样的节点之间完全相连的线路，或者信息之间互为因果的模型，如果要采用通常的线路或者函数来模拟，就需要多得多的线路或者函数才有可能表达。他们以此来论证解释什么是信息整合度和信息整合理论的合理性。但是，他们似乎停止于此。我们没有读到过他们探讨这样的线路的进一步含义的文献，也没有看到他们讨论这样的线路的潜在规则和后果，以及对这样的线路的用途的设想。

\section{叠加体和解叠器} 
我们在了解PQR线路后，深感这里面隐含若干非常重要的东西，并且疑惑这些东西是否可以帮助我们。我们是这样看的：如果把PQR线路分拆开，它就退化成3个非常简单的布尔线路，没有任何独特，当然是在图灵计算的范围。但是，如果把这些非常简单的布尔线路叠加起来，情况就非常不同了。叠加使得P，Q，R都既是函数值又是变量值，也即：因即是果，果即是因。如果再看作常规的计算，就必然会有很多困惑。正如Toroni等人指出的那样，如果硬要用经典计算来模拟PQR线路，就需要多得多的线路。这也就是说，一个仅用3个节点的PQR线路就需要几十个节点的前馈线路才能模拟。正好，我们一直以来在积极思考怎么超越计算复杂性。因此，仅用3个节点就可以做很多事情的PQR线路就促使我们思考，是否把若干布尔线路叠加，就可能创造出一种机会，使得超越计算复杂性成为可能。因此，我们在这个方向上进行了长期的思考和探索。

基于这样的思考，我们提出一个由叠加的布尔线路形成的模型。布尔线路等价于布尔函数，布尔线路的叠加也就等价于布尔函数的叠加。因为布尔函数更容易叙述，我们就采用布尔函数来讨论。我们首先给出叠加（superposition），叠加体（superpositioner）和复馈（reentry）的定义。注意，在本文中，$B$代表1维布尔空间（二值变量空间），$B^N$代表$N$维布尔空间。

\begin{definition}[\bf 叠加和叠加体]
假设有$N$个比特：$T=\{b_1, b_2, \ldots, b_N\}$，再假设有$N$个布尔函数\\ $\{f_1, f_2, \ldots, f_N\}$，$f_i: B^N \to B, i=1, \ldots, N$，如下面的形式：
\begin{align*}
& b_1 = f_1 (b_1,b_2, \ldots, b_N)      \\
& b_2 = f_2 (b_1,b_2, \ldots, b_N)      \\   
& \ldots \ldots \\ 
& b_N = f_N (b_1,b_2, \ldots, b_N)  
\end{align*}
我们则称函数$f_1, \ldots, f_N$在节点$b_1,\ldots, b_N$处叠加，并且，我们称$f_1, \ldots, f_N$和$b_1,\ldots, b_N$形成一个叠加体，称$b_1,\ldots, b_N$为叠加体的节点，称$f_1, \ldots, f_N$为叠加体的复馈函数，称$f_i$复馈到$b_i$处，称$N$为叠加体的尺度。
\hfill$\blacksquare$
\end{definition}
  
注意到，在定义中的单个函数$f_1, \ldots, f_N$，本身没有任何独特之处，就是普通的布尔函数。独特之处在于，这些函数的函数值又复馈给了变量值。这就是叠加，而全体叠加就形成了叠加体。我们现在来看一些叠加体的例。

\begin{example}[\bf 叠加体的例] 
一些最简单的叠加体：\\
{\bf 单点叠加体。}这个叠加体仅有一个比特$b$，因此也就仅有一个布尔函数$f: B \to B$，复馈就是$b = f(b)$。注意到，这样的布尔函数总共有4个：$1, 0, b, \neg b$。选取$f = \neg b$是唯一合理的选择，其他的都没有什么意义。因此我们也称它为自反叠加体。这个叠加体非常简单，但我们后面会看到，即使是这样的极简单的叠加体，也可以起很重要的作用。

{\bf 阴阳叠加体。}这个叠加体仅有两个节点，两个复馈函数。节点为：$b_1, b_2$，复馈函数为：
$$b_1 = f_1(b_1, b_2) = b_1 \land b_2, \ \ \ b_2 = f_2(b_1, b_2) = b1 \oplus b2$$
同样的，还是两个节点，如果安排不同的复馈函数，还可以形成不同的叠加体，如阳阳叠加体，阴阴叠加体，负阴阳叠加体，等等。这些叠加体一样有两个节点，两个复馈函数，但是它们的复馈函数有所不同。例如阳阳叠加体的复馈函数为：
$$b_1 = f_1(b_1, b_2) = b_1 \land b_2, \ \ \ b_2 = f_2(b_1, b_2) = b1 \land b2$$
而负阴阳叠加体的复馈函数为：
$$b_1 = f_1(b_1, b_2) = b_1 \lor b_2, \ \ \ b_2 = f_2(b_1, b_2) = b1 \oplus b2$$
后面我们主要考虑阴阳叠加体。

{\bf PQR叠加体。}前面所列的PQR线路，就是一个简单叠加体，有3个节点和3个复馈函数：~\eqref{eq:p}，~\eqref{eq:q}，~\eqref{eq:r}。我们称这个叠加体为PQR叠加体。

{\bf 五行叠加体。}这个叠加体有5个节点：$b_1, b_2, b_3, b_4, b_5$，5个复馈函数：
\begin{align*}
& b_1 = sk(b_1, b_4, b_5)      \\
& b_2 = sk(b_2, b_5, b_1)      \\   
& b_3 = sk(b_3, b_1, b_2)      \\
& b_4 = sk(b_4, b_2, b_3)      \\
& b_5 = sk(b_5, b_3, b_4)  
\end{align*}
此处$sk$是一个布尔函数，$sk(x, y, z) = (y \land \neg z) \lor (x \land \neg (y \oplus z))$，我们称$sk$为生克函数，其意义很明显：$y$生$x$，$z$克$x$。如果熟悉传统的五行，可以清楚看到为什么我们把这个叠加体称为五行叠加体。
\end{example}

叠加体的一个最基本特征就是叠加，即函数值复馈给变量，然后形成新的函数值，而且可以这样一直下去。为了帮助我们看得更清楚，以PQR叠加体为例。我们假设，在节点P，Q，R处赋值为$(1,1,1)$。如果我们不考虑复馈函数的影响，这时在这些节点处取值，我们就会仍然得到值：$(1,1,1)$。但是，函数值是复馈到节点的，节点处的值就将受到复馈函数的影响。因为是复馈，即自变量的值是函数值，函数值也是自变量的值，因此复馈函数的影响可以是多重的。不仅如此，在在考虑复馈函数时，复馈的次序也很重要。例如，如果我们先应用函数$f_1$，再应用函数$f_3$，然后在节点处取值，就会得到值：$(1,1,0)$； 而如果我们先应用函数$f_3$，再应用函数$f_2$，再应用函数$f_1$，然后在节点处取值，就会得到值：$(0,0,0)$。 这就是说，如果我们要在节点取值，就必须要考虑复馈函数的次序和次数。这就造成了这样一种现象：只有在清楚了应用复馈函数的次数和次序后，谈论节点处的值才是有意义的。但是，在PQR线路中，对次数和次序完全没有任何限制，也就是说任意次数和次序都是允许的。因为叠加，节点处的值首尾相连：在被一个函数赋值后，又成为另一个函数的变量值，而且这个过程重复不停，就形成了值的变动，值的震荡，直到取值的那个时刻。因此，叠加体中节点处的值，除非完全确定了应用各个复馈函数的次数和顺序，就是不完全确定的。这是非常有意义的基本性质。基于这样的思考，我们来定义复馈构型（reentry configuration）。

\begin{definition}[\bf 复馈构型和内蕴函数]
假设叠加体$C$由节点$T=\{b_1, b_2, \ldots, b_N\}$和复馈函数\\ $\{f_1, f_2, \ldots, f_N\}$组成。一个复馈函数$f_j: B^N \to B, j=1, \ldots, N$，可以形成一个$N$维布尔空间（由节点组成）的映射$F_j: B^N \to B^N$，构成如下： 
$$F_j(b_1, b_2, \ldots, b_N) = (b_1,  \ldots, f_j(b_1, b_2, \ldots, b_N), \ldots, b_N)$$
就是说，其它分量都不动，仅$j$分量由复馈函数$f_j$来取值。可以看到$F_j$由$f_j$唯一决定，因此我们可以把$F_j$记为$f_j$。因此在这样的理解下，$f_j$既是复馈函数，也是一个$N$维布尔空间的映射。这无歧义。另外，为方便起见，我们定义$F_0$为$N$维布尔空间的恒等映射：
$$F_0: B^N \to B^N, F_0(b_1, b_2, \ldots, b_N) = (b_1, b_2, \ldots, b_N)$$
而且按照同样的意义，可以把$F_0$记为$f_0$。
对应于一个正整数序列序列：$S=(j_1,\ldots,j_K)$，$j_1,\ldots$\\
是1到$N$的数，$K$是整数，大小没有限制，可以形成一个$N$维布尔空间的映射：
$$h: B^N \to B^N, h(b_1, b_2, \ldots, b_N) = f_{j_1}\ldots f_{j_K}(b_1, b_2, \ldots, b_N)$$
这里的$f_j: B^N \to B^N$，如上理解，是一个$N$维布尔空间的映射。我们称这样的序列$S$为一个复馈构型，称对应的$N$维布尔空间的映射$h$为复馈构型$S$所形成的映射，记为$h = C \diamond S$。另外还有一个特殊的复馈构型为$S=(0)$，这个复馈构型对应的映射为如上的$f_0$。
注意到，一个映射$h$是由$N$个布尔函数构成的，即:
$$h(b_1, b_2, \ldots, b_N) = (h_1(b_1, \ldots, b_N), \ldots, h_N(b_1, \ldots, b_N))$$
我们称这些布尔函数$h_1, \ldots, h_N$为复馈构型$S$所形成的布尔函数，我们也称这些函数为叠加体$C$的内蕴函数。
\hfill$\blacksquare$
\end{definition}

很明显，自变量本身就是函数，根据如上定义，自变量函数，即如$b_1, b_2, \ldots$这样的函数也都是内蕴函数。后面还将更多讨论内蕴函数。有很多复馈构型，因此，虽然一个叠加体的复馈函数仅有$N$个，但是叠加体中可能有远比$N$多的内蕴函数。不同的复馈构型形成的函数可能是相同的。事实上，复馈构型有无穷多个，但是，内蕴函数只能有有限多个。我们来考察一些叠加体的内蕴函数的例。


\begin{example}[\bf 单点叠加体] 
这个叠加体仅有一个节点$b$，一个复馈函数$f$，$b = f(b) = \neg b$。这是最简单的叠加体。复馈构型也很简单，就是这样的序列：$(1), (1,1), (1,1,1), \ldots$。如果序列是奇数长度，则形成的内蕴函数为$\neg b$；如果是偶数长度，则形成的内蕴函数为$b$。节点处（仅一个节点）的取值和复馈构型的对应，可以看表1：
\begin{table}[h]
\centering
    \begin{tabular}{|c|c|c|c|c|c|c|c|c|c|c|c|c|c|c|c|c|}
        \hline
       $b$     & $(1)$ & $(1,1)$ & $(1,1,1)$ & $(1,1,1,1)$ & $(1,1,1,1,1)$  \\ 			\hline
       1      & 0     & 1     & 0     & 1     & 0                \\       \hline
       0      & 1     & 0     & 1     & 0     & 1                \\       \hline
    \end{tabular}
    \caption{单点叠加体-复馈构型和取值}
\end{table}
\end{example} 

就是说，单点叠加体的内蕴函数集为：$\{b, \neg b\}$，仅有两个成员。再看阴阳叠加体。

\begin{example}[\bf 阴阳叠加体] 
这个叠加体有两个节点$b_1, b_2$，复馈函数为：$b_1 = b_1 \land b_2, \ b_2 = b1 \oplus b2$。节点处的取值和复馈序列的对应，可以看表2：
\end{example}
\begin{table}[h]
\centering
    \begin{tabular}{|c|c|c|c|c|c|c|c|c|c|c|c|c|c|c|c|c|}
        \hline
       $b_1, b_2$     & $(1)$ & $(2)$ & $(2,1)$ & $(1,2)$ & $(1,2,1)$  & $(2,1,2)$	  \\              \hline
       $1, 1$       & $1, 1$     & $1,0$    & $1,0$     & $0,0$     &     $0,0$    & $0,0$    \\                 \hline
       $1, 0$       & $0, 0$     & $1,1$    & $0,0$     & $1,1$     &     $0,0$    & $1,0$    \\                 \hline
       $0, 1$       & $0, 1$     & $0,1$    & $0,1$     & $0,1$     &     $0,1$    & $0,1$    \\                 \hline
       $0, 0$       & $0, 0$     & $0,0$    & $0,0$     & $0,0$     &     $0,0$    & $0,0$     \\                 \hline
    \end{tabular}
    \caption{阴阳叠加体-复馈构型和取值}
\end{table}
 
可以看到，在上表中继续增加复馈序列的长度，并不会带来更多的内蕴函数了。因此，阴阳叠加体的内蕴函数集为：$\{b_1, b_1\land b_2, b_1 \land \neg b_2, 0, b_2, b_1\oplus b_2, \neg b_1 \land b_2\}$，总共7个。注意到，叠加体仅有2个复馈函数，但是内蕴函数共有7个，而全部两个变量的布尔函数共有16个。注意到，在内蕴函数集中，前面4个是在$b_1$处取值的内蕴函数，后面3个是在$b_2$处取值的内蕴函数。

注意到，在阴阳叠加体中，复馈函数''$\land$''，起的作用是阳，即集中和聚集的作用；而复馈函数“$\oplus$”，起的作用是阴，即平衡和调和的作用，因此称这个叠加体为阴阳叠加体。很容易看到，如果两个复馈函数都是阳（''$\land$''，所谓阳阳叠加体），则叠加体的行为就简单很多了，其内蕴函数仅有4个。如果两个复馈函数都是阴（阴阴叠加体），叠加体的行为也简单。我们再看PQR叠加体。

\begin{example}[\bf PQR叠加体] 
这个叠加体有3个节点$b_1, b_2, b_3$（P，Q，R），复馈函数为：$b_1 = f_1 = b_2 \lor b_3, \ b_2 = f_2 = b_3, \ b_3 = f_3 = b1 \oplus b2$。节点处的取值和复馈序列的对应，可以看表3：
\end{example} 
\begin{table}[h!t]
\centering
    \begin{tabular}{|c|c|c|c|c|c|c|c|c|c|c|c|c|c|c|c|c|}
        \hline
       $b_1, b_2, b_3$  & $(1)$ & $(2)$ & $(3)$ & $(1,3)$ & $(3,2)$ & $(1,2,3)$  & $(3,1)$	&  \ldots  \\              \hline
       $1, 1, 1$   & $1, 1, 1$  & $1, 1, 1$   & $1, 1, 0$   & $1, 1, 0$   & $1, 1, 0$   &   $0, 0, 0$    &  $1, 1, 0$   &  \\     \hline
       $1, 1, 0$   & $1, 1, 0$  & $1, 0, 0$   & $1, 1, 0$   & $1, 1, 0$   & $1, 0, 1$   &   $0, 0, 0$    &  $1, 1, 0$   &  \\     \hline
       $1, 0, 1$   & $1, 0, 1$  & $1, 1, 1$   & $1, 0, 1$   & $1, 0, 1$   & $1, 1, 0$   &   $1, 1, 1$    &  $1, 0, 1$   &  \\     \hline
       $1, 0, 0$   & $0, 0, 0$  & $1 ,0, 0$   & $1, 0, 1$   & $1, 0, 1$   & $1, 0, 1$   &   $1, 1, 1$    &  $0, 0, 0$   &  \\     \hline
       $0, 1, 1$   & $1, 1, 1$  & $0, 1, 1$   & $0, 1, 1$   & $1, 1, 1$   & $0, 1, 1$   &   $1, 1, 1$    &  $1, 1, 0$   &  \\     \hline
       $0, 1, 0$   & $1, 1, 0$  & $0, 0, 0$   & $0, 1, 1$   & $1, 1, 1$   & $0, 0, 0$   &   $1, 1, 1$    &  $1, 1, 0$   &  \\     \hline
       $0, 0, 1$   & $1, 0, 1$  & $0, 1, 1$   & $0, 0, 0$   & $0, 0, 0$   & $0, 1, 1$   &   $0, 0, 0$    &  $1, 0, 1$   &  \\     \hline
       $0, 0, 0$   & $0, 0, 0$  & $0, 0, 0$   & $0, 0, 0$   & $0, 0, 0$   & $0, 0, 0$   &   $0, 0, 0$    &  $0, 0, 0$   &  \\     \hline
    \end{tabular}
    \caption{PQR叠加体-复馈构型和取值}
\end{table}


PQR叠加体的复馈序列很多，在表3中，我们仅列出了一些，还有很多复馈序列和相应的取值没有列出。可以看一个例，$C$为PQR叠加体，$S = (1, 3)$是一个复馈构型，则对应于$S$的映射就是：$f_{1}f_{3}$，不难知道这个映射是$C \diamond S = (b_1 \lor b_2, b_2, b_1 \oplus b_2)$，这里有3个布尔函数：$b_1 \lor b_2$，$b_2$，和$b_1 \oplus b_2$，它们都是PQR叠加体的内蕴函数。再如，$S = (3, 2)$是另一个复馈构型，则$C \diamond S = (b_1, b_3, b_1 \oplus b_3)$，因此，$b_3, b_1 \oplus b_3$也都是PQR叠加体的内蕴函数。又一方面，如果考虑节点$b_1$，则在这个节点的取值函数起码有这些：$b_1, b_2 \lor b_3, b_2 \lor (b_1\oplus b_2), \ldots$。

前面这3个例给了我们很多启发，揭示了叠加体的内蕴函数的丰富性，也揭示了叠加体的非逻辑性和不确定性。我们现在来考虑怎么对待这种非逻辑性和不确定性。

首先看叠加体的非逻辑性和不确定性。如果仅看定义3.1中的复馈函数，没有任何非逻辑性和不确定性。问题出在复馈，以及因此而形成的复馈构型。叠加体中，复馈函数的输出同时也是输入（对照前馈线路），而且叠加体中的输入和输出并没有清晰的分界（对照循环线路），这是叠加体的根本特性。就是说，如果不强行施加一个复馈构型到叠加体，叠加体本身是含混的；如果把一个复馈构型强加于叠加体，叠加体就退化成为普通的$N$维布尔空间的映射。但是，复馈构型并不是唯一的，一个叠加体有很多复馈构型，其中的任何一个复馈构型都不具备优越性和优先性，这些复馈构型都是可能的。即使是像单点叠加体这样简单的叠加体，也是如此。而正是复馈构型的非唯一性带来了叠加体的不确定性。这个不确定性不是来自于外部，而是内部结构必然形成的。因此，我们应该认为在叠加体中，所有的复馈构型形成的内蕴函数都同时存在，它们形成的取值也同时存在，这是非逻辑的。再说一遍，叠加体内部的复馈函数遵循严格的逻辑，可以严格计算，但是，因为复馈，形成了非逻辑成分，而且是不可去除的非逻辑成分。如果要清理非逻辑成分，就需要强行施加一个复馈构型，但是又因为复馈构型的不唯一性，就导入了非确定性。这种非逻辑性和不确定性，是叠加体的根本性质。这些性质的确是非经典的，非逻辑的，但是可以说非常有用的，我们正是要探索怎么最充分利用这些性质。

叠加体的这种非逻辑性和不确定性，导致叠加体中节点的取值的不确定性。如果我们关注叠加体的某个节点，我们就可以看到这个节点处的取值并不确定，只有当复馈序列确定后，这个取值才能确定。而且这个取值将随复馈序列的变动而变动。例如，对于单点叠加体，仅有一个节点，那么当复馈序列是一个奇数序列时，取值就是$\neg b$，当复馈序列是一个偶数序列时，取值就是$b$。就是说，节点的取值，是在0和1之间变动，并不能确定。只有当复馈序列确定时，这个值才得到确定。这样的叠加体，这样的性质，正是许多应用所需要的，也是在应对很多情况时应该形成的。我们后面可以看到，单点叠加体虽然简单，但正好可以用来产生/描述那种内在的犹豫。即使是最简单的叠加体，也可以起很重要的作用。

再例如阴阳叠加体。如果考虑节点$b_1$，则所有的在此节点的取值函数为：$b_1, b_1\land b_2, b_1 \land \neg b_2, 0$；
如果考虑节点$b_2$，则所有的在此节点的取值函数为：$b_2, b_1\oplus b_2, \neg b_1 \land b_2$。就是说，在节点$b_1$处，共有4个取值函数，而在节点$b_2$处，共有3个取值函数。这是对所有复馈序列，无论复馈序列的长度。就是说，仅有这些内蕴函数。

因此，我们看到，叠加体和前馈函数（前馈线路）完全不同。前馈函数（线路）中的作用是单向的，因此也就是确定的，因和果是清晰分开的。但是，叠加体则不同，因为有了复馈，因和果也就失去了清晰的分界。这样的后果是，在叠加体中，很多函数叠加在一起，这在经典计算中根本无法表达、无法描述。面对这样的一种情况，我们采用的一种方式是导入复馈序列，在把一个复馈序列强加于叠加体后，叠加体退化成了经典情况，这时，经典计算就可以进行。然而，在叠加体中复馈序列是很多的，而且任一复馈序列都没有任何优先。

但是，这是在叠加体内才如此。在叠加体外部，完全可以对叠加体内的复馈序列产生倾向，而且可以利用这种倾向和叠加体的配合来完成某些任务。这就是我们将要讨论的解叠（disposition），和能够做解叠的工具，即解叠器（dispositioner）。

\begin{definition}[\bf 解叠和解叠器]
假设叠加体$C$由节点$T=\{b_1, b_2, \ldots, b_N\}$，和复馈函数$\{f_1, f_2, \ldots, f_N\}$\\ 组成，给定一个复馈构型：$S=(j_1,\ldots,j_K)$，就得到一个$N$维布尔映射$h: B^N \to B^N, h = C \diamond S =f_{j_1}\ldots f_{j_K}$，而这个映射的$j_1$分量，就是一个内蕴函数，即$g: B^N \to B, g = (C \diamond S)_{j_1}$。这个内蕴函数在$b_{j_1}$处取值，而且这个取值是完全确定的。在此意义上，就是说，$S$解叠了$C$，使得$C$从叠加状态解除，而退化成一个固定的布尔函数。这个过程，从一个复馈构型得到一个布尔函数的过程就称为解叠。

因为复馈构型并不唯一，选取复馈构型完全可以带有倾向。解叠器就是这样的一个装置，使得可以按照某种倾向来选取复馈构型，并且依据复馈构型做解叠。
\hfill$\blacksquare$
\end{definition}

在解叠时，因为复馈构型确定了，叠加体就退化成了普通的布尔空间的映射，因此也就成为了经典的图灵计算。但是，当复馈构型不确定时，叠加体+解叠器就不是一个或者几个布尔函数能包括的。这种叠加体+解叠器事实上形成了一种新形态的信息处理方式。来看一些例。

\begin{example}[\bf 自反叠加体的解叠] 
单点叠加体（自反叠加体）仅有一个节点。如前所述，这个叠加体的全部内蕴函数就是：$\{b, \neg b\}$。因此一个单点叠加体的解叠器就是一个以某种倾向选取$b$或者$\neg b$的选择器，或者说，单点叠加体+解叠器，就形成了一个随机装置，在$b$或者$\neg b$之间用一定的概率变动，这种随机装置是信息处理的非常基本的组件。关键在于，这个单点叠加体+解叠器是内生的，而非外来的。由此可见，即使是单点叠加体这样非常简单的叠加体，都可以起重要作用。
\end{example}

\begin{example}[\bf 阴阳叠加体的解叠] 
阴阳叠加体有2个节点。如前所述，这个叠加体的全部内蕴函数共有7个：$\{b_1, b_1\land b_2, b_1 \land \neg b_2, 0, b_2, b_1\oplus b_2, \neg b_1 \land b_2\}$。这些函数中，前面4个是在$b_1$处取值的内蕴函数，后面3个是在$b_2$处取值的内蕴函数。考虑在$b_1$处取值。这样，就有4个函数相关联。也就是说，从函数$b_1$可以关联到函数$0$。这是令人吃惊的，但是事实。因此可以说，这个叠加体中蕴含了这样的情况，即把$b_1$和$0$内在关联了起来。其他的函数，也可以形成类似的关联情况。而且，这样的关联还可以很容易实现，即对复馈序列做一些调整，就可能从一个函数滑动到另一个函数。这些性质是非常可贵的，值得仔细加以利用。

阴阳叠加体的解叠器有可能使得我们利用这些性质。自然的，阴阳叠加体+解叠器也可以如前一个例子那样，用一定的概率来选取内蕴函数，那样也是一种利用叠加体的用途。
\end{example} 

仅从这两个非常简单的叠加体，就可以看到，叠加体+解叠器已经构成了一种新型的信息处理方式，一种新型的计算模型，非常不同于经典计算。

\section{思绪空间及内容}

叠加体，内蕴函数，解叠器都需要存在于某个容器中，只有这样，它们才能配合起来进行合适的计算，也只有在容器中，才能对它们进行各种操作。这个容器就是思绪空间（conceiving space）。

\begin{definition}[\bf 思绪空间]
思绪空间是一个容器，其中有若干比特，和作用于这些比特的布尔函数，思绪空间是让这些比特和布尔函数得以在其中活动的工作空间。
\hfill$\blacksquare$
\end{definition}

我们是在若干年前导入思绪空间的\cite{ulearning}。在这篇文章前，我们认为思绪空间中的主要元素是前馈网络和前馈函数。在导入了叠加体之后，现在我们清楚了，思绪空间可以和叠加体良好结合和协调，叠加体需要思绪空间作为存在的容器，而思绪空间也需要叠加体作为内容，或许，思绪空间最重要的成分就是叠加体，其它的成分都是围绕叠加体的。思绪空间中完全可能有多个叠加体，它们之间可能进行复杂的相互作用。有了叠加体，思绪空间就更加丰富和饱满起来。

我们在前面已经看到，虽然一个叠加体的复馈函数仅有$N$个，但是可能有非常多的内蕴函数。我们这里把内蕴函数的定义再陈述一次，把内蕴函数突出出来。

\begin{definition}[\bf 内蕴函数]
假设$C$是一个尺度为$N$的叠加体，节点为$T=\{b_1, b_2, \ldots, b_N\}$。则其所有的复馈构型形成复馈构型集$W$，注意$W$为一无限集合。对应于任何一个复馈构型$S \in W$，假设$S = (j_1, \ldots, j_K), \ j_i \in [1, N], i = 1, \ldots, K$，$K$是一个整数，不限大小，那么就有一个函数$g: B^N \to B, g = (C \diamond S)_{j_1}$，这个函数在节点$b_{j_1}$处取值，这个函数就是一个$C$的内蕴函数，称为在节点$b_{j_1}$处取值的内蕴函数，也称它由复馈构型$S$形成。另外，特别地，全部自变量函数，即$b_1, b_2, \ldots, b_N$这样的函数也都是内蕴函数。全体内蕴函数就形成$C$的内蕴函数集。
\hfill$\blacksquare$
\end{definition}

注意到，完全可能，不同的复馈构型形成同一个布尔函数$g$（可能取值的节点不同，但仍然是相同的函数）。也完全可能，同一个布尔函数$g$，在不同的节点处取值。再注意到，如果固定一个节点，例如$b_1$，那么所有结尾于1的复馈构型，如$(1, 2), (1, 3, 2), (1, 1, 4, 3)$等等，都将形成一个在$b_1$处取值的内蕴函数。这些内蕴函数叠加在节点$b_1$上。

叠加体的全体内蕴函数都是思绪空间的成员，也就是说，如果思绪空间中有了一个叠加体，也就会有了很多内蕴函数作为成员。可以很容易看到，正是因为内蕴函数，叠加体成为了思绪空间的核心。围绕一个叠加体，非常多的布尔函数可以形成，就有了很大的布尔函数群。如果没有叠加体，就非常难以表达这样规模的布尔函数群。对一个叠加体$C$而言，究竟$C$内蕴了那些布尔函数？在$C$的一个节点$b_j$处，究竟叠加了多少布尔函数？这些问题非常重要，而且绝对不平凡，它们反映了叠加体的基本性质，决定叠加体的功能。

注意到，我们在定义内蕴函数的时候，仅用到了节点，并没有考虑节点处的值的来源。我们知道，思绪空间中有很多布尔函数，如果我们在叠加体的每个节点处都连接上一个布尔函数，然后让复馈构型起作用，这样就会形成了一个布尔函数。这就是用叠加体来建构布尔函数。这是叠加体的一个重要功能。我们有下面的定义。

\begin{definition}[\bf 在叠加体上构建的布尔函数]
假设叠加体$C$由节点$T=\{b_1, b_2, \ldots, b_N\}$，和复馈函数$\{f_1, f_2, \ldots, f_N\}$组成。假定有一组布尔函数$g_i: B^M \to B, i = 1, \ldots, N$，我们把$g_i$连接到$b_i, i=1, \ldots, N$上，即在$b_i$处的取值是$g_i(x_1, \ldots, x_M)$。对任何一个复馈构型：$S=(j_1,\ldots,j_K)$，我们得到一个内蕴函数$g: B^N \to B, \  g = (C \diamond S)_{j_1} $，注意$g$的自变量为$b_1, \ldots, b_N$，即$g(b_1, \ldots, b_N) = (C \diamond S)_{j_1}(b_1, \ldots, b_N)$，这时我们用$g_i(x_1, \ldots, x_M), i=1, \ldots, N$来取代$b_i, i = 1, \ldots, N$，我们就得到一个布尔函数：
$$h: B^M \to B, \  h = (C \diamond S)_{j_1}(g_1(x_1, \ldots, x_M), \ldots, g_N(x_1, \ldots, x_M))$$
我们称$g_1, \ldots, g_N$为馈入函数，函数$h$为在叠加体$C$上由复馈序列$S$和馈入函数构建的布尔函数，我们把这个函数记为：
$h = C \diamond S \diamond (g_1, \ldots, g_N)$。注意到，$M$不同于$N$，通常是$M$远大于$N$。
\hfill$\blacksquare$
\end{definition}

可以看到，$h = (C \diamond S)_{j_1}(g_1(x_1, \ldots, x_M)$，就是一个内蕴函数但是其自变量处代入馈入函数。容易看到，用这样的方式，通过一个叠加体，可以构建非常多的布尔函数。这是思绪空间中构建布尔函数的一种普遍方式。这些所有的构建的布尔函数，都通过叠加体联系起来了。可以看例子。

\begin{example}[\bf 构建布尔函数] 
先考虑自反叠加体。假设$f: B^M \to B$是任何布尔函数，则通过自反叠加体构建出$f$和$\neg f$。这样，这两个函数就通过叠加体联系起来了。而且通过解叠器，还可以对这两个布尔函数形成一定概率的选择（内生的犹豫），以及滑动到另一个（变化到反面）。

对阴阳叠加体，因为阴阳叠加体的内蕴函数为：$\{b_1, b_1\land b_2, b_1 \land \neg b_2, 0, b_2, b_1\oplus b_2, \neg b_1 \land b_2\}$，那么很容易构建这些布尔函数：假设$f, g: B^M \to B$是两个布尔函数，则$\{f, f \land g, f \land \neg g, 0, g, f \oplus g, \neg f \land g\}$都是在阴阳叠加体上构建出来的布尔函数。这些布尔函数都是通过阴阳叠加体联系起来的，可以很容易在这些函数之间形成概率选取和滑动。
\end{example} 

当然例中的构建的函数都很简单，但是，作为例子，可以帮助我们看清楚叠加体和构建的函数的作用。当叠加体的尺度增加，复杂度增加，构建出来的函数就会非常丰富，而且这些函数通过叠加体联系起来，而且也可以通过解叠器获得选取和相互间的滑动。因此，通过前面的定义在思绪空间中构建函数，是很重要的工具。这样构建出来的布尔函数，和带参数的布尔函数有密切关系。我们先看带参数的布尔函数的定义。

\begin{definition}[\bf 带参数的布尔函数]
一个布尔函数$\varphi: B^N \times B^J \to B$，我们称为在$B^N$上带有$J$个参数（二值变量参数）的布尔函数. 我们通常把这个函数写成$\varphi(x, p): B^N \to B, \ x \in B^N, p \in B^J$. 
\hfill$\blacksquare$
\end{definition}

\begin{example}[\bf 带参数的布尔函数]
$f: B^2 \to B, f(x_1, x_2) = x_1 \land x_2$是一个最简单的布尔函数，但是这个函数不带参数。我们来看一个和$f$相关，但是带参数的布尔函数：$\varphi(x_1, x_2, s) = x_1 \land (x_2 \oplus s)$。可以很容易看到，当$s = 0$，我们有$\varphi(x_1, x_2, 0) = f(x_1, x_2) = x_1 \land x_2$，当$s=1$，则$\varphi(x_1, x_2, 1) = f(x_1, x_2) = x_1 \land \neg x_2$。这就是说，参数不同时，我们获得两个不同的函数。 
\end{example} 

注意到，前面例中的带参数的布尔函数，可以包括在阴阳叠加体构建出来的布尔函数中。具体是这样的：当参数$s=0$时，$\varphi$成为了$b_1 \land b_2$，这就是当复馈构型为$S=(1)$时得到的内蕴函数；当参数$s=1$时，$\varphi$成为了$b_1 \land \neg b_2$，这就是当复馈构型为$S=(1, 2)$时得到的内蕴函数。就是说，这个带参数的布尔函数所包括的布尔函数，其实都可以在阴阳叠加体上构建出来。我们也可以称，这个带参数的布尔函数可以由阴阳叠加体完全表达出来。事实上，我们将看到这是普遍情况，即任何带参数的布尔函数，都可以找到一个叠加体来完全表达。下面的引理就来证明此点。

\begin{lemma}[\bf 完全表达]
假设$\varphi(x, p): B^M \to B, \ x \in B^M, p \in B^J$是一个带参数的布尔函数，则存在一个叠加体$C$，以及存在一组馈入函数$g_i: B^M \to B, i = 1, \ldots, N$，使得，对任何$p \in B^J$，都可以找到一个复馈构型$S_p$，使得$\varphi(x, p) =  C \diamond S_p \diamond (g_1, \ldots, g_N)$。
\end{lemma}
{\bf 证明：}本引理没有对叠加体做任何限制，因此，叠加体$C$的存在，和复馈函数的存在是明显的。我们仅需要让$N=2^J$，并且对集合$B^J$做一个排序：$p_i \in B^J, i = 1, \ldots, N=2^J$，然后对每一个$p \in B^J$都选取$g_i: B^M \to B, g_i(x) = \varphi(x, p_i),  i = 1, \ldots, N=2^J$，这样，当然有$\varphi(x, p_i) =  C \diamond S_{p_i} \diamond (g_1, \ldots, g_N)$，此处$S_{p_i} = (0)$。
\hfill$\blacksquare$

当然，证明仅是很简单的说明存在，说明完全表达可以做到。理想中的用于构建带参数的布尔函数的叠加体，应该用最小的$N$来做到。这就是一个困难的问题，应该留待后续的讨论。

这个引理说明，对任何一个带参数的布尔函数，我们都可以找到一个叠加体，一组馈入函数，使得对每一个参数，使用这个参数的布尔函数都表达为一个叠加体上构建的布尔函数。这就是说，我们把一组布尔函数，表达成了由叠加体来构建的内容。因为叠加体可以完全表达任何带参数的布尔函数，因此叠加体就可以比带参数的布尔函数表达更多的内容，而且是多得非常多。更重要的是，带参数的布尔函数是外界施加的，很难对其做些什么。而对叠加体构建的内容，则可以做很多事情，可以通过复馈序列滑动和选取的。此点很关键。还有一个非常关键的区别：在带参数的布尔函数中，不同的布尔函数是分开的，但是，在叠加体构建出来的布尔函数中，不同的布尔函数是叠加在一起的。叠加在一起将起非常重要的作用。

带参数的布尔函数在计算中可以起重要作用。我们在\cite{partition}中定义了一种用于带参数的布尔函数上的算子，我们称为试错算子，用这个算子就可以把非常难以表达的问题表达得非常清晰和简洁。由此可见带参数的布尔函数的重要作用。事实上，对一个计算特例，如果事先知道参数，往往可以形成指数级别的跨越。怎么可能事先知道参数呢？这就是大问题。事实上，图灵的神谕机就那种可以事先知道参数机器，带参数的布尔函数和图灵的神谕机有非常密切的关系。

对于神谕机，可以把它看成两个部分，一个部分为图灵机T，一个部分为神谕O。因为O都从T的外部施加到T的，T不能对O做任何事情，只能遵循而执行。但是，对于叠加体则不同，叠加体构建的函数，是从内部构建的，完全可以根据经验，根据结果，来加以调整，因此，在此意义上，叠加体要强于神谕机。而且，可能更重要的还在于：即使是神谕，不同的参数形成的函数仍然是分开的，孤立的，而叠加体构建的函数是叠加在一起的。

\begin{center}
**********
\end{center}

给定一个叠加体，就会有相应的内蕴函数集。例如自反叠加体的内蕴函数集有两个成员，阴阳叠加集的内蕴函数集有7个成员。内蕴函数集的内部也会有独特的有趣的结构。这些都由叠加体的复馈函数所决定的。但是，复馈函数怎么决定这个结构？而这个结构又是什么样的？我们相信，这些问题将对我们理解叠加体起重要作用。但是，目前我们还无法深入探讨这些问题，我们将持续思考。			

这样，我们就可以看到思绪空间中的丰富内容。首先，思绪空间中有大量的二进制值节点，连接节点的有大量的布尔函数。这些布尔函数可以组成前馈网络，也可能形成循环网络，更有趣的是，这些布尔函数可以形成复馈，就产生了叠加体。而因为有了叠加体，就可以构建更多的函数，也可以形成带参数的函数。还有解叠器，以及解叠器形成的解叠。这样思绪空间就更加丰富了。我们认为，在这样的思绪空间中，一些经典计算很难实现的计算将可能得以实现（见下节）。

\section{应用和实现}
我们相信，叠加体+解叠器作为一种新型的信息处理模型和计算模型将发挥重大的作用。现在我们来设想一些可能的叠加体+解叠器的用途。

\textbf{\textit{编程和学习}} \\ 
通过前面定义的叠加体上构建的布尔函数，我们事实上形成了一种新的编程方式和学习方式。假设有一个叠加体$C$，如果我们选择馈入函数$g_i: B^M \to B, i=1, \ldots, N$，然后再选择复馈构型$S = (j_1, \ldots, j_K)$，从而得到内蕴函数$f = C \diamond S \diamond (g_1, \ldots, g_N)$。完全可以把这个函数$f$的过程看作一种编程，即通过选择馈入函数和复馈构型来实现的，这是一种新型的编程。如果我们是采用数据驱动的方法来获取馈入函数和复馈构型，也可以把这个过程看成学习。这是一种新型的学习。

这种用途，即固定复馈构型，把叠加体退化成为了一个固定的布尔函数。这是一个非常强有力的形成新的布尔函数的方式，代价很低，仅需要选取馈入函数和复馈构型。可以对照其他编程方法和机器学习方法。如果要人工编程，就需要把一个布尔函数用逻辑门表达出来，然后写入程序；如果通过机器学习，就需要很大的一个数据集（合适采样集），然后用机器学习的方法，用这些数据来形成这个布尔函数。相对于这些，基于叠加体来进行编程和学习，就容易了很多。事实上，一个叠加体本身已经包括了某种倾向，自然可以使得符合这种倾向的编程和学习变得容易。这种编程方式和学习方式是消耗很低的，可以用于非常迅速的编程或学习。是否有可能在人脑或者动物脑中，广泛存在基于叠加体的编程方式和学习方式？这个思路值得高度重视。

\textbf{\textit{内生感受}} \\ 
参看例3.5，就可以看到自反叠加体可以用于产生一种简单但是极其重要的内生感受，即犹豫。对更多更复杂的内生感受，就需要更复杂的叠加体。在智能体内部形成各种内生感受，对智能体的主观有很重要的作用。

\textbf{\textit{类比和联想}} \\ 
我们希望在智能体内部形成类比和联想，而且我们还希望可以触及它们，控制它们，并且利用它们。经典计算并不提供普遍的方法来形成类比和联想，乃至控制和利用类比和联想。从对叠加体的初步讨论中，我们看到叠加体的潜在能力。类比和联想似乎就在叠加体中。以阴阳叠加体为例，看例3.6，在那里，在节点$b_1$处取值的内蕴函数有4个，其中包括$b_1 \land b_2$和$0$。这两个函数非常不同，但是，因为它们通过叠加体联系到了一起，如果滑动复馈序列，也就滑动了内蕴函数，而滑动复馈序列是非常容易的。而且，这两个函数还是叠加在一起的。这就和我们心目中的类比和联想很接近了。因此，假如我们有一些对象，要达到这些对象的类比和联想，我们就需要有某个合适的叠加体，使得这个叠加体的内蕴函数和那些对象联系起来（具体怎么联系，则完全可能变化多样）。当做到这样之后，对象之间的类比就成为了两个内蕴函数之间的滑动，对象之间的联想就成为了从一个内蕴函数到另一个内蕴函数的跳动。这是经典计算难以支持的，但是叠加体+解叠器却可以很容易支持。

类比和联想，特别是怎么在人工智能体中实现它们，是困扰了人工智能和认知科学很久的课题。现在有了叠加体，让我们看到了一种新的境界，这非常有意义。

\textbf{\textit{形成理解}} \\ 
究竟什么是智能体对某个东西形成了理解？理解又是什么？对这些问题，我们这里用叠加体来提供一种独特的视角。通常，某个东西在智能体中会获得某种表征，而且可以对这种表征做信息处理。但是，即使能对这种表征做出很完善正确的信息处理，也很难说智能体理解了这个东西。这是因为，很可能因为某种原因，智能体内部的表征有了微小变动，因此信息处理就出现错误，那样，就不能说智能体理解了。在通常的意义中，对某个东西达到理解是要求比较高的。但是，我们来设想这样的情况：即在智能体内对某个东西的表征可以达到了某个叠加体中，这个叠加体对和这个东西相关的信息有融合，然后，在后续的信息处理中，并不是这些表征在直接起作用，而是叠加体中的内蕴函数在起作用，那么，对这个东西的信息处理，就会周延很多。这和我们心目中对理解的看法就接近了。

什么是理解，也是困扰了人工智能和认知科学很久的课题。用叠加体，我们应该可以把对智能体的理解的认识向前推进一步。

\textbf{\textit{产生能动}} \\ 
智能体的能动就是：智能体可以超越预设的程序，而采取新的行动。如果一个智能体里面完全是经典计算，恐怕这个智能体不可能产生能动，因为在经典计算中，一起都是预设好的。但是，在叠加体中，情况非常不一样。叠加体中，所有的内蕴函数是叠加在一起的，因此，所有的内蕴函数也是可能的，而且内蕴函数的数量可能非常巨大。这就为能动形成了可能性。当然，究竟能动是如何实现的，还有待进一步的工作，但是叠加体允许能动，而经典计算排斥能动，则是清晰的。

如前面所述，正是因为追求智能体中的能动，我们才来到了叠加体的前面。

\textbf{\textit{参与主观的形成}} \\
智能体的主观是其内部的倾向，各种倾向，如前面说的内生感受，犹豫等，这种倾向可能有很多。应该说，没有叠加体，完全采用经典计算，也是可以形成倾向的。但是，叠加体可以极大帮助。例如，对一个叠加体，我们知道一组内蕴函数叠加在某一些节点上，这就形成了一种倾向，即这些节点的取值只能由这一组内蕴函数来做。而这种倾向，用经典计算来体现，就相当困难。

除开设想一些分别的用途之外，我们还应该设想这些用途在智能体中的联合运用。非常可能，一个智能体中，需要若干个叠加体+解叠器联合工作才能达到良好的效果。思绪空间中，非常可能有若干体器，有的体器用于形成动机，有的用于产生内生感受，有的用于产生类比和联想，有的用于形成能动，有的作为执行部分，有的用于新型编程，等等。这些体器配合工作，可以形成一个能力很强的智能体。这是我们的基本思路。

\begin{center}
**********
\end{center}

前面讨论了叠加体的潜在应用，现在我们对物理实现叠加体做一些最简单的讨论。叠加体和解叠器都是数学对象，自然是首先做数学讨论，然后才考虑物理实现。我们已经知道，叠加体和解叠器不是经典计算可以完善描述的，但是我们知道脑中有叠加体（正是来自于脑科学的工作刺激我们形成叠加体概念的），而且脑中的叠加体在信息处理中发挥重要作用，因此，我们有理由充分相信可以做到叠加体的物理实现。实现叠加体和解叠器的途径有两种，一是用硬件，一是用软件。

先来看软件途径。现在的计算软件都是在经典计算框架内，因此，不可能用软件来完全实现任意的叠加体和解叠器。但是，可以用软件在一定程度上模拟叠加体和解叠器。这里的关键是叠加体的尺度。如果增加叠加体的尺度/增加复馈函数的复杂度，其内蕴函数就会非常多。这时要用随机变量来模拟全部内蕴函数，就成为事实上不可能。因此，实际上模拟的随机变量就只能覆盖一些内蕴函数（而且数量很小）。这是可以做的。然而，软件模拟的局限性就凸显了出来。叠加体内部包含非常丰富的内蕴函数，但是软件模拟仅能选择少量的来模拟。选什么呢？这一个困境。这个困境就是经典计算的困境。这也是为什么叠加体+解叠器超越经典计算。

再看硬件途径。叠加体本身是非逻辑的，这个非逻辑性使得不可能用现有的硬件方式来实现完全的叠加体，因为现有的硬件方式仍然是经典计算范围内的。叠加体要求其内蕴函数可以同时被计算出来，并且叠加在一起，目前还没有一种硬件可以做到。这就需要我们去探索和利用各种物理性质，去发明新的物理器件，来具体实现叠加体。但是，有一点可以明确，那就是，人脑的能力说明，这样的硬件是存在的，我们需要把它们发明出来，造出来。当新的硬件出现后，与之匹配的软件也会出现，那时就可以在新的水平上软硬齐发来实现叠加体。

正如脑科学研究者发现的那样，在人脑和动物脑中都有复馈，因此也就有叠加体。在脑中，叠加体一定是有力的信息处理装置，否则进化不会让这样的装置存在下去。另一方面，研究叠加体将使得信息处理的能力更上层楼。因此应该把脑科学和意识科学的成果，和对叠加体的研究结合起来，这样可以极大促进这两个方向的发展。

\section{一些评论}
{\bf 1.} 前面的讨论表明，叠加体+解叠器形成的信息处理，具备比图灵计算更强大的信息处理能力。如果不计资源的消耗，图灵计算也可以在某种程度上模拟实现叠加体+解叠器所形成的信息处理。但是，那仅是模拟，并不能完全复现。而且，这种模拟要消耗高得多的资源。因此，我们应该把叠加体和解叠器形成的计算结合进计算的框架，那样将极大提高计算能力。

{\bf 2.} 图灵很早（就在提出图灵机后的两三年内）就提出了神谕机。神谕机模型可以解释很多计算上的难题。但是，神谕机却是外在于图灵机的，这样在计算模型内部，神谕机不可捉摸。而采用叠加体+解叠器，带参数的布尔函数（神谕机）在模型内部自然产生，因此带参数的布尔函数是可以在模型内部发展，改进，进化的，这就带来了很重要的能力。

{\bf 3.} 智能体可以仅由图灵计算来形成，即智能体内部的计算模块全是图灵计算程序。事实上，迄今为止的智能体都是如此的。然而，如果智能体中具备一个或更多的叠加体，就可以形成更强的能力。我们提出了思绪空间的概念，思绪空间中包含叠加体。完全可以设想，对某项困难计算任务（如求解SAT，求解NPP，等等），如果想要超越计算复杂性，就不能采用图灵计算程序（或者说一个算法）来处理，而应该采用一个具备特别能力的智能体来处理，这个智能体中有思绪空间，其中有一个或者多个由叠加体，它们能形成针对该计算任务的良好的主观和积极的能动，进而使得这个智能体可以对这项计算任务超越计算复杂性。

{\bf 4.} 由叠加体+解叠器形成的计算模型，非常不同于图灵计算。图灵计算是确定性的（当然也可以模拟非确定性的），而叠加体本质上说非确定性的。两者有根本的不同。因为复馈函数的应用的不确定性，叠加体必然内蕴随机。但是，叠加体+解叠器计算又非常不同于传统的从外部加入随机因素。首先，这样的计算的随机是由复馈函数的结构决定了的，因此也有明确的结构。其次，它有学习的成分，可以通过学习来改进。因此，这样的计算有这样的性质：有随机性，但是也有明确的方向性，而且方向是可以通过学习调整的。传统的从外部加入的随机因素很难做到这些。    

{\bf 5.} 叠加体+解叠器和量子计算看起来有非常紧密的联系，从这两种计算的各个方面进行考察，都可以感受到这种联系。在这个方向上探索，应该非常有意义，我们后面将报告我们在这个方向上的思考。

{\bf 6.} 叠加体具备很丰富的内涵。在本文中，我们采用复馈构型去获取这些内涵。但可以肯定，复馈构型并没有穷尽叠加体的内涵，一定还有很多复馈构型不能触及的内涵。我们认为，复馈构型是第一阶的工具，我们应该首先在这上面工作，我们期待在复馈构型带来成功后再回来探索更高阶的工具和内涵。

{\bf 7.} 本文中，我们仅在布尔函数的基础上讨论了叠加体。布尔函数仅是一种计算模型中的一种。完全可以相信，在其他的计算模型上也可以建立叠加体，也就是说，用现代的编程语言来建立叠加体是可能的。当然这仅是一种模拟式的表达，因为叠加体是非逻辑的，表达就不是那种可以由计算机具体执行的表达，仅是可以部分模拟实现的。

{\bf 8.} 叠加体作为整体是经典逻辑不能解释和描述的，是经典计算不能处理的，但是叠加体里面的部分，即复馈函数，却是完全符合经典逻辑和经典计算的。但是，如果我们强行从外部施加一个复馈构型，则叠加体退化成为经典逻辑。如此看，其实就提供了一种思路：经典逻辑是否就是从从外部对复杂事物强加了一个构型？经典逻辑是人的思考工具和认识工具，是否绝对高于一切的？根据叠加体的能力来猜想，很可能叠加体是另外一种思考工具，另外一种认识工具，和经典逻辑相辅相成 。中国文化中的阴阳思考，五行思考，或许就是应用叠加体（阴阳叠加体和五行叠加体）的范例。这些是非常有趣的思考线索。

{\bf 9.} 基于对超越计算复杂性的追求，并且受到意识科学的启发，我们提出叠加体模型。这个模型现在是作为数学模型存在，尚未形成物理的模型，虽然我们相信可以建立物理形态的叠加体。然而，我们还需要进一步工作，展示这个模型的确可以用于超越计算复杂性，至少是理论上如此。也还需要进一步工作，说明为什么这个模型可以实现类别与联想，理解，能动等功能的原理。那样一来，这个模型的意义就得到彰显。


\end{CJK}
\end{document}